# Charge Transfer Induced Molecular Hole Doping into Thin Film of Metal–Organic-Frameworks

**Deok Yeon Lee[†], Eun-Kyung Kim[†], Nabeen K. Shrestha[†]\*, Danil W. Boukhvalov[†]\*, Joong Kee Lee[‡], Sung-Hwan Han[†]**

[†]Department of Chemistry, Hanyang University, Seoul 133-791, Republic of Korea.
[‡]Energy Storage Research Center, Korea Institute of Science and Technology, Seoul-136-791, Republic of Korea.

\*nabeenkshrestha@hotmail.com; danil@hanyang.ac.kr; shhan@hanyang.ac.kr

**ABSTRACT**

Despite the highly porous nature with significantly large surface area, metal organic frameworks (MOFs) can be hardly used in electronic, and optoelectronic devices due to their extremely poor electrical conductivity. Therefore, the study of MOF thin films that require electron transport or conductivity in combination with the everlasting porosity is highly desirable. In the present work, thin films of $Co_3(NDC)_3DMF_4$ MOFs with improved electronic conductivity are synthesized using layer-by-layer and doctor blade coating techniques followed by iodine doping. The as-prepared and doped films are characterized using FE-SEM, EDX, UV/Visible spectroscopy, XPS, current-voltage measurement, photoluminescence spectroscopy, cyclic voltammetry, and incident photon to current efficiency measurements. In addition, the electronic and semiconductor property of the MOF films are characterized using Hall Effect measurement, which reveals that in contrast to the insulator behavior of the as-prepared MOFs, the iodine doped MOFs behave as a *p-type* semiconductor. This is caused by charge transfer induced hole doping into the frameworks. The observed charge transfer induced hole doping phenomenon is also confirmed by calculating the densities of states of the as-prepared and iodine doped MOFs based on density functional theory. Photoluminescence spectroscopy demonstrate an efficient interfacial charge transfer between $TiO_2$ and iodine doped MOFs, which can be applied to harvest solar radiations.

# 1. INTRODUCTION

Metal–organic-frameworks (MOFs) have been studied extensively for many promising applications such as gas storage, gas decontamination/separation, heterogeneous catalysis, photocatalysis, sensors, and electrical energy storage.[1-9] For most of these applications, MOFs are used in the form of bulk powder. One of the interesting and challenging future issues in the field of MOFs is their material investigation for making thin film, and study their electronic properties.[10-15] However, huge majority of MOFs are electrically insulators.[16-21] Therefore, despite the highly porous nature with significantly large surface area, MOFs can be hardly used in electronic, and optoelectronic devices. Hence, the study of MOF thin films that require electron transport or conductivity in combination with the everlasting porosity is highly desirable. Recently, the concepts of MOFs as potentially active materials in semiconductor science are emerging.[22-26] However, the most challenging issue in MOFs is their poor electrical conductivity, which is needed to be improved to use MOF thin film as a semiconductor material. Some attempts to improve conductivity demonstrate that the conductivity of MOFs can be improved to some extent by molecular doping of the frameworks with electron acceptors, or by using some typical organic linkers.[16-21,27] In addition, modulation of MOF conductivity by changing the metal cations has also been demostrated. For instance, Park *et al.*[28] have recently demonstrated that the shortest S⋯S distance between neighboring tetrathiafulvalene cores of tetrathiafulvalene tetrabenzoate-based MOFs can be shorten by using larger cations. This causes to increase the overlaping between the S $3p_z$ orbitals, and thereby modulates the electrical conductivity of the MOFs from $Zn^{2+}$ to $Co^{2+}$, $Mn^{2+}$, and $Cd^{2+}$. On the other hand, Sun *et al.*[29] have demonstrated that by changing cations from $Mn^{2+}$ to $Fe^{2+}$ in dihydroxybenzene-1,4-dicarboxylate-based MOFs, electrical conductivity can be enhanced to a million-fold.

In post-synthetic modification route of electrical conductivity, iodine has been often inserted into MOFs as electron acceptor molecule, which has demonstrated an improved conductivity of the frameworks as a result of oxidative doping of the MOFs by guest $I_2$.[18,19] The oxidative doping of the framework system is commonly due to the oxidation of metal ion of the frameworks.[18] However, the oxidative doping of MOFs is not straight forward if the oxidation of metal ion of the frameworks is not feasible thermodynamically. In such systems, oxidative doping of the frameworks could be due to guest $I_2$-ligand π electron interaction. Although, host-guest chemistry of MOF-$I_2$ systems, and an improved electrical conductivity of the frameworks due to host $I_2$ molecules have been reported[19], detail mechanism behind the oxidative doping of the frameworks has not been studied yet. In addition, after the oxidative doping, advantage on application of such MOFs with improved conductivity has been hardly investigated. In the present work, a thin film of Co-based MOFs, *viz.* cobalt (II) 2,6-naphthalendicarboxylic acid (i.e., *Co₃(NDC)₃DMF₄*, which is denoted here as *Co₃(NDC)₃*), is investigated as a model MOFs. The MOF films are deposited onto a nonconducting amine functionalized glass substrate using layer-by-layer (LbL) and Doctor-Blade (DB) techniques. After iodine doping, the framework films are investigated for their electronic and semiconductor properties. The current work demonstrates that doping of iodine into the frameworks significantly improve the electrical conductivity within the frameworks. The detail mechanism behind the improved conductivity of the iodine doped frameworks has been studied experimentally, and the observed results have been confirmed using computational chemistry (DFT calculation) of the system. Further, as a concept-of-demonstration for application, it has been demonstrated that a facile transfer of photo-generated electron from LUMO level of iodine doped *Co₃(NDC)₃* to conduction band of ITO or $TiO_2$ takes palace, and the observed interfacial charge transfer phenomenon can be employed to harvest solar radiations.

## 2. Experimental Section

MOF thin films were prepared using doctor blade and layer-by-layer techques. Details on the deposition procedures, iodine doping, film characterizations and DFT calculations are given in Supporting Information section.

## 3. RESULTS AND DISCUSSION

Scheme S1 shows the progress of *Co₃(NDC)₃* thin film formation using LBL and DB coating techniques. SEM images of the *Co₃(NDC)₃* film on a glass substrate using DB (A), and LbL (B) coating techniques are shown in Figures 1, which exhibit similar rod like morphology of the frameworks in both films. However, the cross-sectional views of the films show that the LBL film has compact layers, while the DB film has microprous structure with large voids. The composition of the films determined by EDX is shown in Figure S1 (Supporting Information). The XRD patterns of the bulk and the thin film of *Co₃(NDC)₃* frameworks are shown in Figure 2. The diffraction patterns clearly show that the bulk *Co₃(NDC)₃* synthesized by the standard solvothermal process[30] and the *Co₃(NDC)₃* film grown by LbL and DB techniques have the same characteristic crystalline structure. This demonstrates the successful synthesis of thin MOF films.

*Co₃(NDC)₃* films were further characterized using UV/Visible absorption spectroscopy. *Co₃(NDC)₃* framework layers on a glass slide shows a characteristic absorption peak at about 548 nm in Figure 3 (A) and Figure S2 (A), which is due to the d-d transition (i.e., $t_{2g}$ (d7) to $e_g$ transitions) associated with the $Co^{II}$ centers of the frameworks. However, in addition to this peak, both the films after iodine doping exhibited a new peak at about 438 nm. This peak position is exactly matching with the peak of iodine solution (Figure S2 (B)). Therefore, the peak at about 438 nm can be ascribed to the characteristic absorption peak due to iodine doping. Additionally, XPS was employed to investigate the doping of iodine into the frameworks. The XPS survey spectra of the as-prepared and doped films are shown in Figure S3 (A). The doped MOFs show I 3d XPS peaks (Figures S3 (A) suggesting the capturing of iodine molecules inside the porous frameworks. Quantitatively, 0.35 molecules of iodine is found to be accommodated into 1 unit cell of *Co₃(NDC)₃* frameworks. As reported previously,[19] capturing of iodine from iodine-acetonitrile solution was also noticed visually as well as by UV/visible absorption measurement (Figure S4). Figure 3 (B) shows the electrical conductivity of the MOF films deposited on a glass substrate measured in terms of current passing through the film under an external voltage bias. Originally, the as-prepared MOF films without doping show the insulating behavior. However, regardless of the methods of film formation, the framework films after iodine doping allow the passage of currents under external bias. As compared to the DB film, the LBL film exhibited little higher conductivity which could be due to lower number of grain boundary present in the LBL film. It will be noteworthy here to discuss briefly that the electronic absorption peak and the absorption edge positions of the *Co₃(NDC)₃* film before and after iodine doping are the same in Figure 3(A). Based on this data, the *Co₃(NDC)₃* film before and after doping should have the similar energy gap and electronic property. Therefore, the apparent insulating behavior of the undoped films could be due to extremely poor conductivity as a result of poor carrier concentration (supposed to be virtually zero carriers) in the undoped materials. Further, the electronic and semiconducting properties of the *Co₃(NDC)₃* films were characterized using a Hall Effect measuring device. When the MOF films on a non-conducting glass was investigated, they exhibit Hall Effect with the electronic conductivity in the range of $10^{-6}$ (LBL film) ~ $10^{-7}$ s cm$^{-1}$ (DB film), the bulk charge carrier concentration in the range of ~ +5 × $10^{11}$ cm$^3$, and the Hall coefficient is in the range of ~ +1 × $10^7$ m$^2$ C$^{-1}$ (Table 1, Table S1, and Figure S5 (A) - (B)). As shown in Table 1 and Table S1, the conductivity of the LBL film is found to be little higher than the DB film. In addition, when the undoped MOF films were investigated for the Hall Effect, the instrument displayed "*connection fail*" message in the operating description window of the device monitor (Fig. S5 (C)), which indicates that the electrical circuit for the measurement could not be established. This finding again suggests the insulating behavior of the undoped MOF films. These types of conductivity behaviors are in line with the current *vs* applied bias shown in Figure 3 (B). In addition to the electronic property, it should be noted in Table 1, Table S1, and Fig. S5 (A)-(B) that the bulk-charge carrier concentration and the Hall coefficient exhibited by the doped MOF films are positive, which implies that the doped MOFs are behaving as a *p-type* semiconductor. These results suggest that before iodine doping, the charge carrier concentration in the frameworks is virtually zero. However, iodine doping is working here as oxidative doping and therefore, creating holes as charge carriers in the frameworks. The oxidation of the MOFs by iodine is also supported by the XPS detection of $I_3^-$ (Figure S3(B)) in the iodine doped frameworks.

Each sub-building block unit of *$Co_3(NDC)_3$* consists of linear arrangement of three Co(II) ions linked via six carboxylate groups of the NDC ligands (Figure S6). The crosslinking of these trinuclear sub-building units via the naphthalene rings forms 1-D channels along *a*- and *c*-axis, and thereby, it produces a neutral 3-D network structure of the frameworks. The detail structures of the *$Co_3(NDC)_3$* can be found elsewhere.[30,31] The frameworks have shown dominating antiferromagnetic interactions between Co(II) ions of the frameworks perticularly in the high-temperature region, and a noticeable ferrimagnetic behavior in the low temperature region. The detail magnetic structure and property of the frameworks can be found in the reference [31]. During iodine doping, the iodine guest molecules are expected to be confined to the network structure of the frameworks surrounded by aromatic rings. Because of the intermolecular interactions between iodine guest molecules and π-electrons from the aromatic ring of the host, such arrangements can result in cooperative charge transfer. Therefore, the observed conductivity of the iodine doped film can be considered as a result of interaction between iodine and the aromatic ring of the frameworks. The oxidization of $Co^{2+}$ ion of the frameworks by iodine is thermodynamically not feasible due to their unfaborable redox potentials. This argument is also supported experimentally by XPS spectra of the frameworks before and after iodine doping in Figure S3(C), where no shift in binding energy positions of the Co 2p peaks before and after iodine doping can be observed. Instead of metal ions, probably the charge transfer interaction between iodine and ligand particularly with π-electrons from the aromatic ring of the ligands takes place, which leads to the oxidation of the frameworks, and thereby hole doping into the frameworks takes place. To confirm the above charge transfer reaction leading to oxidation of the aromatic ring, detailed studies on C 1s XPS spectrum of the frameworks were made. In contrast to the undoped frameworks, the deconvoluted C 1s peaks of the iodine doped frameworks show a small new peak at 285.5 eV, which is located at higher binding energy position than the peak due to C-C bond in Figure S3(D). This new peak can be ascribed to the consequence of charge transfer interaction between iodine and the aromatic ring of the ligands leading to the above readox reaction.[32] Further, Figure 4(A) shows the UV/Visible absorption spectra of free $I_2$, and $Co_3(NDC)_3$ before and after iodine doping. A similar UV/Visible spectrum can also be observed in the case of iodine doped naphthalene dycarboxylic acid (Figure 4(B)). In addition to the characteristic peak from iodine at about 512.8 nm, a new peak in deep UV region (i.e., at about 241.5 nm) can be observed. This additional absorption peak of the doped materials can be assigned to the donor-acceptor charge transfer complex formation[33,34] (i.e., π electrons from aromatic ring donor and $I_2$ acceptor interaction). These findings clarifies the above observed hole doping characteristic of the iodine doped $Co_3(NDC)_3$ films in Hall Effect measurements.

In order to further confirm the charge transfer induced hole doping into frameworks by iodine doping, quantitative estimation of the charge transfer were performed using *ab initio* calculations based on DFT (for details, see Supporting Information SI-1). Results of the calculation are shown in Figure 5, which demonstrate that after iodine doping into $Co_3(NDC)_3$ MOFs, the HOMO level of the frameworks remain unchanged, and iodine levels appear over it (Figure 5(A)). In contrast, the LUMO level of the frameworks after iodine doping shifts down to about 0.2 eV, which is attributed to the charge transfer from $Co_3(NDC)_3$ to $I_2$, and hole doping of the frameworks. Accepting electrons from $Co_3(NDC)_3$, or naphthalene dicarboxylic acid molecules also provides almost similar energy level shift of the $I_2$ levels from the values of a pure $I_2$ molecule (Figure 5(B)). The observed shift is in qualitative agreement with the appearance of charge transfer complex formation in Figure 4 after iodine doping into $Co_3(NDC)_3$, and naphthalene dicarboxylic acid molecules. Note that small quantitative difference in the shift of levels in the case of doping on naphthalene and $Co_3(NDC)_3$ can be observed, which is actually caused by the smearing of the energy levels of iodine on $Co_3(NDC)_3$ because, in contrast to naphthalene, the frameworks have bulk-like electronic structure. This is concluded based on the almost identical value of transferred electrons and binding energies in both substrates. A similar calculated shift of the energy levels of iodine molecule has been reported to be in quantitative agreement with the shift calculated by quantum chemical methods after $I_2$ adsorption on benzene molecule.[35] The distnce between iodine molecule and naphthalene rings (see Figure S6) is 3.28 Å for the case of adsorption on naphthalene dicarboxylic acid and 3.22 Å for adsorption on *Co_3(NDC)_3* MOFs which is smaller than typical van der Waals bonds (over 3.5 Å). Calculated value of binding energies for adsorption of $I_2$ on *Co_3(NDC)_3* MOFs and naphthalene dicarboxylic acid is about 0.25 eV/$I_2$. This value is in order higher than typical values of van der Waals bonds (below 0.02 eV) due to enhancement of bond by charge transfer. On the other hand, the calculated value is smaller than typical value of hydrogen bonds in water (0.43 eV/$H_2O$). So, we can conclude that at room temperature only part of the doped iodine molecules is adsorbed forming charge transfer complex with *Co_3(NDC)_3* MOFs or naphthalene dicarboxylic acid. This conclusion is in good agreement with the results of measurements of I 3d XPS spectrum (Figure S3(B)), which reveals that only one from about 6.7 iodine molecules adsorbed on MOFs with charge transfer. The total number of electrons transferred from MOFs or naphthalene to iodine molecule is about 0.02 electrons per $I_2$. Using the above data, we can estimate the number of holes induced in $cm^3$ of the MOFs by charge transfer complex with iodine as follows: (i) The volume of unit cell of *Co_3(NDC)_3* MOFs is 5,000 $Å^3$.[29] So, there are $2 \times 10^{20}$ unit cells per $cm^3$. (ii) Experimental results demonstrate that there is about one $I_2$ per 3 unit cells (note that unit cell contains 3 atoms of cobalt). Thus, there are about $6.6 \times 10^{19}$ $I_2$/$cm^3$. (iii) The concentration ratio of $I_3^-$ (which is participated in the formation of charge transfer bonds) and free $I_2$ is about 1:6.7, and thus the number of charge transfer per $cm^3$ of the frameworks is about $9.8 \times 10^{18}$. (iv) To calculate the total number of holes induced by charge transfer between MOFs and $I_2$, we multiply the obtained number by calculated value of transferred electrons (i.e., 0.02 electrons per $I_2$), and thus finally we obtain $1.9 \times 10^{17}$ holes/$cm^3$ of the *Co_3(NDC)_3* MOFs. The above calculated hole concentration is in several orders higher than the experimental value obtained by Hall Effect measurement. The difference is due to the fact that we describe here rather idealistic situation where all charge carriers are assumed to participate without any barriers from grain boundaries. However, in actual practice, existence of lots of grain boundaries, and other structural imperfection in MOF film resist electron transfer for the charge transfer complex formation. Current theory of mobility in 1D systems[36] demonstrates that its value depends on the elastisity of lattice and effective mass of charge carriers, which, on the other hand, depend on the shapes of band structure. Adsorption of molecules does not affect elastic properties of MOF but can provides some changes in band structure. More prospective source of the manipulation of mobility in doped MOF is the varying of atomic structure of MOF that can provide dramatic changes in its band structure.

The semiconducting property of the MOF films is further characterized by measuring their HOMO-LUMO positions. Using the absorption spectra of the $Co_3(NDC)_3$ film, the optical band gap of iodine doped frameworks is estimated to be about 2.2 eV (Figure S7). The experimentally estimated energy gap of the doped frameworks is close to the calculated (about 2.1 eV) value. From the onset of ground state oxidation peak in the cyclic voltammogram, the HOMO energy level is estimated to be -6.2 eV *vs* vacuum (Figure S8). Based on the energy gap and the HOMO energy level, the LUMO energy level of the $Co_3(NDC)_3$ frameworks is estimated to be about -3.9 eV. Recently, Butler *et al.*[37] have estimated HOMO-LUMO positions of a number of MOFs using DFT theory. However, the measured HOMO-LUMO energy levels of $Co_3(NDC)_3$ of the present work is different from those of other MOFs reported by them. As pointed out by them, some differences could arise from different methods used to measure the ionozation potentail. It shold be noted that the measured HOMO-LUMO energy levels of $Co_3(NDC)_3$ are closer to those of other transitional metal besed- MOFs measured by the same cyclic voltammetric method.[38-40] Considering the strong visible light absorption characteristic of the frameworks as exhibited by Figure 3(A) and the suitable HOMO-LUMO positions, an interfacial electron transfer from LUMO level of the frameworks to a *n-type* material with lower conduction band such as ITO (-4.7 eV) or $TiO_2$ (-4.2 eV) could be established, and this phenomenon can be applied to light harvesting devices. To demonstrate this, a thin $Co_3(NDC)_3$ film was layered onto an ITO glass substrate, and $TiO_2$ film coated on a non-conducting glass substrate. The interfacial electron transfer phenomenon from the $Co_3(NDC)_3$ frameworks to the ITO or $TiO_2$ was studied using PL spectroscopy.

Figure 6 shows intense photoluminescence $Co(II)^*$ emission spectra of the undoped and doped $Co_3(NDC)_3$ films, which have exhibited a significantly higher photoluminescence quenching after iodine doping. The $Co(II)^*$ emission quenching can be attributed to an efficient interfacial electron transfer from LUMO of the frameworks to the conduction band of ITO. Quantitatively, the LBL and DB framework films have exhibited 97.4% and 95.1% of the $Co(II)^*$ emission quenching. A similar efficient interfacial electron transfer quenching phenomenon of the $Co(II)^*$ emission was also observed when the iodine doped $Co_3(NDC)_3$ frameworks are deposited to a $TiO_2$ film. The observed efficient interfacial electron transfer reveals the energy harvesting characteristic of the iodine doped frameworks. However, when a similar experiment was performed by layering a thin film of $Co_3(NDC)_3$ on a non-conducting glass substrate, no photoluminescence quenching was observed (Figure S9). This finding reveals the existence of a facile electron transportation path from cobalt d-d transition of the frameworks to the ITO or $TiO_2$ film only after iodine doping. The successful interfacial charge transfer after iodine doping of the frameworks can be ascribed to the improved conductivity of the frameworks by charge transfer induced hole doping. The conductivity of the $Co_3(NDC)_3$ films after iodine doping is higher by a factor of $10^3$ than that of the iodine[18]. As a result of semiconducting behavior of the iodine doped $Co_3(NDC)_3$ film and its ability to inject photoelectrons to ITO or $TiO_2$, such materials could be used to construct a device that can be used in various fields. As an example of demonstration, the observed interfacial charge transfer phenomenon is employed here to harvest solar radiations. For this, doctor bladed $TiO_2$ mesoporous film on a FTO substrate was sensitized with $Co_3(NDC)_3$ layers using LBL techniques. After iodine doping, a Grätzel type cell with a Pt counter and $I^-/I_3^-$ redox electrolyte was constructed, and the light harvesting capability of the iodine doped $Co_3(NDC)_3$ layer was investigated in terms of the incident photon to current efficiency (IPCE). The IPCE spectrum shown in Figure S10 shows the maximum efficiency at the wavelength of the incident light where the absorption of light is maximum in the UV/Visible absorption spectra (Figure 3(A)). This suggests that the $Co_3(NDC)_3$ film is working here as a light harvesting layer. As compared to the undoped film, the IPCE response in Figure S10 is attributed to the successful interfacial electron injection from the doped $Co_3(NDC)_3$ frameworks to the $TiO_2$ films.

## 4. CONCLUSIONS

In summary, thin films of $Co_3(NDC)_3$ frameworks were fabricated using layer-by-layer and doctor blade coating techniques, and the electrical conductivity of the films was improved significantly by iodine doping. In the current work, the undoped film shows insulating behavior while the doped film exhibits Hall Effect with *p-type* characteristic. The modification from insulator to *p-type* characteristic of the doped films are ascribed to the hole doping into the framework film due to the charge transfer complex formation. The HOMO-LUMO positions of the doped frameworks are found to be suitably situated for the efficient interfacial photoelectron injection from the frameworks to ITO or $TiO_2$. This result shows that the Co-based MOFs can be a potential new semiconducting material. As large number of metal ions and variety of organic linkers are available, an infinite number of combinations are possible to fabricate MOF films with further improved electronic and optical properties, which can have potential application in various fields.

**Supporting Information**

Experimental details on MOF film deposition, iodine doping, film characterization, DFT calculations, and additional results on characterization and application aspect of MOF films.

**ACKNOWLEDGMENTS**

This research was supported by the KIST Institutional Program (2E23964) and by Basic Science Research Program through the National Research Foundation of Korea (NRF) funded by the Ministry of Education (2013009768).

**REFERENCES**

(1) Czaja, A. U.; Trukhan N.; Muller, U. Industrial Applications of Metal–Organic Frameworks. *Chem. Soc. Rev.* **2009**, *38*, 1284-1293.

(2) Choi, K. M.; Jeon, H. J.; Kang, J. K.; Yaghi, O. M. Heterogeneity within Order in Crystals of a Porous Metal–Organic Framework. *J. Am. Chem. Soc.* **2011**, *133*,11920-11923.

(3) Deng, H.; Doonan,C .J.; Furukawa, H.; Ferreira, R. B.; Towne, J.; Knobler, C. B.; Wang, B.; Yaghi, O. M. Multiple Functional Groups of Varying Ratios in Metal-Organic Frameworks. *Science* **2010**, *327*, 846-850.

(4) Silva, C.G.; Corma, A.; Garcia, H. Metal–Organic Frameworks as Semiconductors. *J. Mater. Chem.* **2010**, *20*, 3141-3156.

(5) Silva, C. G.; Luz, I.; Llabres i Xamena, F. X.; Corma, A.; Garcia, H. Water Stable Zr–Benzenedicarboxylate Metal–Organic Frameworks as Photocatalysts for Hydrogen Generation. *Chem.–Eur. J.* **2010**, *16*, 11133-11138.

(6) Wang, D. E.; Deng, K. J.; Lv, K. L.; Wang, C. G.; Wen, L. L.; Li, D. F. Structures, Photoluminescence and Photocatalytic Properties of Three New Metal–Organic Frameworks based on Non-rigid Long Bridges. *Cryst. Eng. Comm.* **2009**, *11*, 1442-1450.

(7) Das, M. C.; Xu, H.; Wang, Z.; Srinivas, G.; Zhou, W.; Yue, Y. F.; Nesterov, V. N.; Qian, G.; Chen, B. A $Zn_4O$-Containing Doubly Interpenetrated Porous Metal–Organic Framework for Photocatalytic Decomposition of Methyl Orange. *Chem. Commun.* **2011**, *47*, 11715-11717.

(8) Lee, D. Y.; Yoon, S. J.; Shrestha, N. K.; Lee, S. -H.; Ahn, H.; Han, S. -H. *Microporous Mesoporous Mater*. **2012**, *153*, 163-165.

(9) Lee, D. Y.; Shinde, D. V.; Kim, E. -K.; Lee, W.; Oh, I. -W.; Shrestha, N. K.; Lee, J. K.; Han, S. -H. *Microporous Mesoporous Mater*. **2013**, *171*, 53-57.

(10) Hermes, S.; Schroder, F.; Chelmowski, R.; Woll, C.; Fischer, R. A. Selective Nucleation and Growth of Metal−Organic Open Framework Thin Films on Patterned $COOH/CF_3$-Terminated Self-Assembled Monolayers on Au (111). *J. Am. Chem. Soc.* **2005**, *127*, 13744-13745.

(11) Hermes, S.; Zacher, D.; Baunemann, A.; Woll, C.; Fischer, R. A. Selective Growth and MOCVD Loading of Small Single Crystals of MOF-5 at Alumina and Silica Surfaces Modified with Organic Self-Assembled Monolayers. *Chem. Mater.* **2007**, *19*, 2168-2173.

(12) Gascon, J.; Aguado, S.; Kapteijn, F. Manufacture of Dense Coatings of $Cu_3(BTC)_2$ (HKUST-1) on [alpha]-Alumina. *Microporous. Mesoporous Mater.* **2008**, *113*, 132-138.

(13) Arnold, M.; Kortunov, P.; Jones, D. J.; Nedellec, Y.; Karger, J.; Caro, J. Oriented Crystallisation on Supports and Anisotropic Mass Transport of the Metal-Organic Framework Manganese Formate. *Eur. J. Inorg. Chem*. **2007**, *1*, 60-64.

(14) Scherb, C.; Schodel, A.; Bein, T. Directing the Structure of Metal–Organic Frameworks by Oriented Surface Growth on an Organic Monolayer. *Angew. Chem., Int. Ed*. **2008**, *120*, 5861-5863.

(15) Lu, G.; Farha, O. K.; Zhang, W.; Huo, F.; Hupp, J. T. Engineering ZIF-8 Thin Films for Hybrid MOF-Based Devices. *Adv. Mater* **2012**, *24*, 3970-3974.

(16)     Gandara, F.; Uribe-Romo, F. J.; Britt, D. K.; Furukawa, H.; Lei, L.; Cheng, R.; Duan, X.; O'Keeffe, M. Yaghi, O. M. Porous, Conductive Metal-Triazolates and Their Structural Elucidation by the Charge-Flipping Method. *Chem. Eur. J.* **2012**, *18*, 10595 – 10601.

(17)     Talin, A. A.; Centrone, A.; Ford, A. C.; Foster, M. E.; Stavila, V.; Haney, P.; Kinney, R. A.; Szalai, V.; Gabaly, F. E.; Yoon, H. P.; Léonard, F.; Allendorf, M. D. Tunable Electrical Conductivity in Metal-Organic Framework Thin-Film Devices. *Science* **2014**, *343*, 66-69.

(18)     Kobayashi, Y.; Jacobs, B.; Allendorf, M. D.; Long, J. R. Conductivity, Doping, and Redox Chemistry of a Microporous Dithiolene-Based Metal−Organic Framework. *Chem. Mater.* **2010**, *22*, 4120–4122.

(19)     Zeng, M. H.; Wang, Q. X.; Tan, Y. X.; Hu, S.; Zhao, H. X.; Long, L. S.; Kurmoo, M. Rigid Pillars and Double Walls in a Porous Metal-Organic Framework: Single-Crystal to Single-Crystal, Controlled Uptake and Release of Iodine and Electrical Conductivity. *J. Am. Chem. Soc.* **2010**, *132*, 2561-2563.

(20)     Hendon, C. H.; Tiana, D.; Walsh, A. Conductive Metal–Organic Frameworks and Networks: Fact or Fantasy? *Phys. Chem. Chem. Phys.* **2012**, *14*, 13120-13132.

(21)     Sun, L.; Miyakai, T.; Seki, S.; Dinca, M. $Mn_2$(2,5-disulfhydrylbenzene-1,4-dicarboxylate): A Microporous Metal–Organic Framework with Infinite $(-Mn–S-)_\infty$ Chains and High Intrinsic Charge Mobility. *J. Am. Chem. Soc.* **2013**, *135*, 8185-8188.

(22)     Rao, K.V.; Datta, K. K. R.; Eswaramoorthy, M.; George, S. J. Light-Harvesting Hybrid Assemblies. *Chem. Eur. J.* **2012**, *18*, 2184-2194.

(23)     Kent, C. A.; Liu, D.; Ma, L.; Papanikolas, J. M.; Meyer, T. J.; Lin, W. Light Harvesting in Microscale Metal Organic Frameworks by Energy Migration and Interfacial Electron Transfer Quenching. *J. Am. Chem. Soc.* **2011**, *133*, 12940-12943.

(24)     Jin, S.; Son, H. J.; Farha, O. K.; Wiederrecht, G. P.; Hupp, J. T. Energy Transfer from Quantum Dots to Metal–Organic Frameworks for Enhanced Light Harvesting. *J. Am. Chem. Soc.* **2013**, *135*, 955-958.

(25) Zhan, W. W.; Kuang, Q.; Zhou, J. Z.; Kong, X. J.; Xie, Z. X.; Zheng, L. S. Semiconductor@Metal–Organic Framework Core–Shell Heterostructures: A Case of ZnO@ZIF-8 Nanorods with Selective Photoelectrochemical Response. *J. Am. Chem. Soc.* **2013**, *135*, 1926-1933.

(26) Gao, J.; Miao, J.; Li, P. Z.; Teng, W. Y.; Yang, L.; Zhao, Yanli.; Liu, B.; Zhang, Q. A p-Type Ti(IV)-based Metal–Organic Framework with Visible-Light Photo-Response. *Chem. Commun.* **2014**, *50*, 3786-3788.

(27) Campbell, M. G.; Sheberla, D.; Liu, S. F.; Swager, T. M.; Dinca, M. $Cu_3$(hexaiminotriphenylene)$_2$: An Electrically Conductive 2D Metal–Organic Framework for Chemiresistive Sensing. *Angew. Chem. Int. Ed*. **2015**, *54*, 4349 –4352.

(28) Park, S. S.; Hontz, E. R.; Sun, L.; Hendon, C. H.; Walsh, A.; Voorhis V. T.; Dinca, M. Cation-Dependent Intrinsic Electrical Conductivity in Isostructural Tetrathiafulvalene-Based Microporous Metal−Organic Frameworks. *J. Am. Chem. Soc.* **2015**, *137*, 1774-1777.

(29) Sun, L.; Hendon, C. H.; Minier, M. A.; Walsh, A.; Dincă, M. Million-Fold Electrical Conductivity Enhancement in $Fe_2$(DEBDC) versus $Mn_2$(DEBDC) (E = S, O). *J. Am. Chem. Soc*. **2015**, *137*, 6164-6167.

(30) Liu, B.; Zou, R. Q.; Zhong, R. Q.; Han, S.; Shioyama, H.; Yamada, T.; Maruta, G.; Takeda, S.; Xu, Q. Microporous Coordination Polymers of Cobalt(II) and Manganese(II) 2,6-Naphthalenedicarboxylate: Preparations, Structures and Gas Sorptive and Magnetic Properties. *Microporous Mesoporous Mater.* **2008**, *111*, 470-477.

(31) Wang, X.-F.; Zhang, Y.-B.; Zhang, W.-X.; Xue, W.; Zhou, H.-L.; Chen, X.-M. Buffering Additive Effect in The Formation of Metal–Carboxylate Frameworks with Slightly Different Linear $M_3(RCOO)_6$ Clusters. *CrystEngComm* **2011**, *13*, 4196–4201.

(32) Zhao, Y.; Wei, J.; Vajtai, R.; Ajayan, P. M.; Barrera E. V. Iodine Doped Carbon Nanotube Cables Exceeding Specific Electrical Conductivity of Metals. *Sci Rep*. 2011, *1*, 83-87.

(33) Refat, M. S.; Grabchev, I.; Chovelon, J. -M.; Ivanova, G. Spectral Properties of New N,N-bis-alkyl-1,4,6,8-Naphthalenediimide Complexes. *Spectrochim. Acta Part A* **2006**, *64*, 435–441.

(34) Camarota, B.; Goto, Y.; Inagaki, S.; Garrone, E.; Onida, B. Electron-Rich Sites at the Surface of Periodic Mesoporous Organosilicas: A UV−Visible Characterization of Adsorbed Iodine. *J. Phys. Chem. C* **2009**, *113*, 20396-20400.

(35) Su, J. T.; Zewail, A. H. Solvation Ultrafast Dynamics of Reactions. 14. Molecular Dynamics and ab Initio Studies of Charge-Transfer Reactions of Iodine in Benzene Clusters. *J. Phys. Chem. A*. **1998**, *102*, 4082-4099.

(36) Beleznay, F. B.; Bogar, F.; Ladik, J. Charge Carrier Mobility in Quasi-one-Dimensional Systems: Application to a Guanine Stack. *J. Chem. Phys.* **2003**, *119*, 5690.

(37) Butler, K.T.; Hendon, C. H.; Walsh, A. Electronic Chemical Potentials of Porous Metal–Organic Frameworks. *J. Am. Chem. Soc.* **2014**, *136*, 2703-2706.

(38) Lee, D. Y.; Shinde, D. V.; Yoon, S. J.; Cho, K. N.; Lee, W.; Shrestha , N. K.; Han, S.-H. Cu-Based Metal–Organic Frameworks for Photovoltaic Application. *J. Phys. Chem. C* 2014, *118*, 16328–16334.

(39) Lee, D. Y.; Shin, C. Y.;Yoon, S. J.; Lee, H. Y.; Lee, W.; Shrestha , N. K.; Lee, J. K.; Han, S.-H. Enhanced Photovoltaic Performance of Cu-based Metal-Organic Frameworks Sensitized Solar Cell by Addition of Carbon Nanotubes. *Sci Rep*. **2014**, *4*,3930-3934.

(40) Lee, D. Y.; Kim, E.-K.; Shin, C. Y.; Shinde, D. V.; Lee, W.; Shrestha, N. K.; Lee, J. K.; S.-H. Han. Layer-by-Layer Deposition and Photovoltaic Property of Ru-based Metal–Organic Frameworks. *RSC Adv*., **2014**, *4*, 12037-12042.

**Table 1**. Various parameters obtained at 25 °C from Hall Effect measurement of $Co_3(NDC)_3$ MOF films prepared by Layer-by-Layer technique on a glass substrate.

| | | | |
|---|---|---|---|
| Bulk concentration | $5.48 \times 10^{11}$ (cm³) | Sheet Concentration | $1.64 \times 10^7$ (cm²) |
| Mobility | 21.2 (cm²/Vs) | Conductivity | $1.88 \times 10^{-6}$ (s/cm⁻¹) |
| Resistivity | $5.37 \times 10^5$ (Ωcm) | Hall Coefficient | $1.14 \times 10^7$ (m²c⁻¹) |

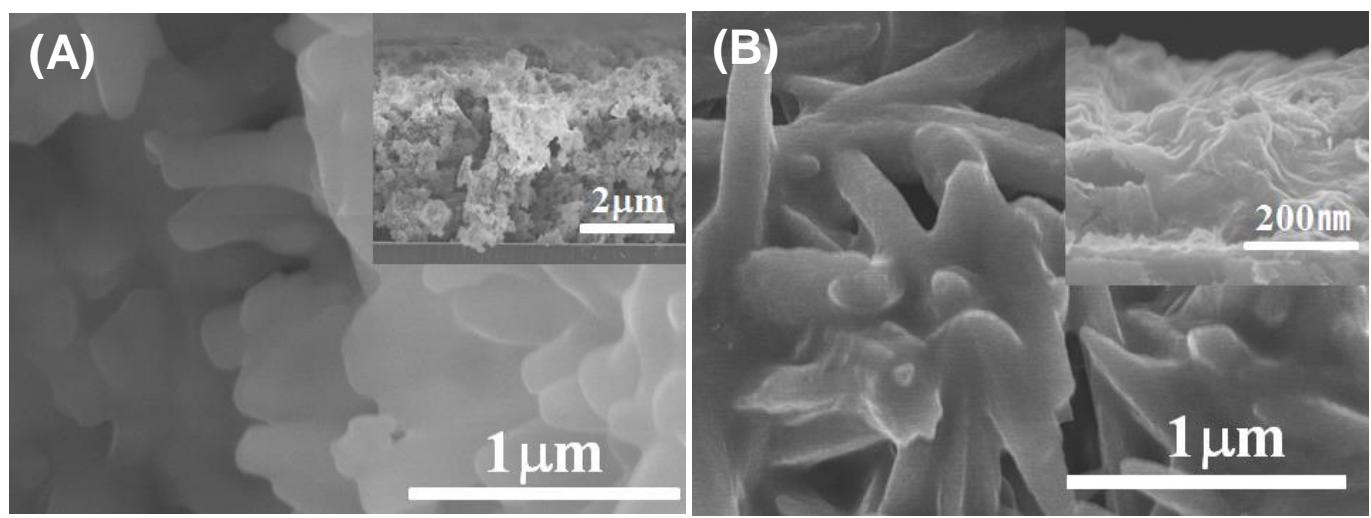

**Figure 1.** Top SEM view of a $Co_3(NDC)_3$ MOFs: Doctor-Blade (A), and LbL (B) films. Insets show the cross-sectional SEM images of the respective films.

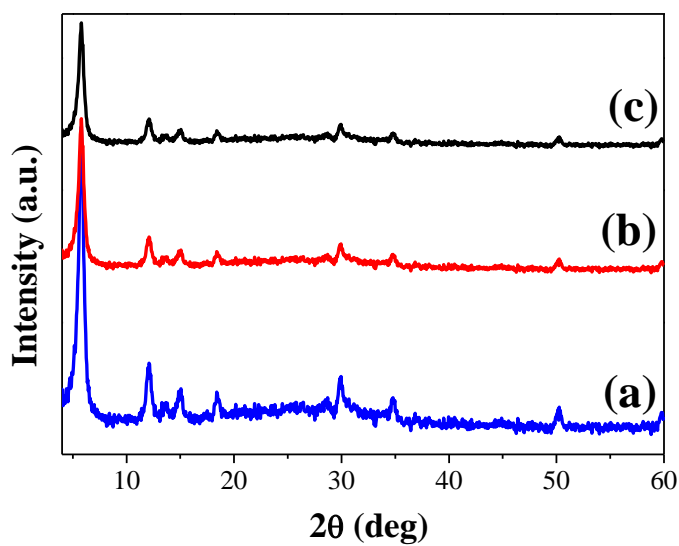

**Figure 2.** XRD patterns of *Co₃(NDC)₃* MOFs: (a) bulk, (b) Doctor-Blade film, and (c) LbL film.

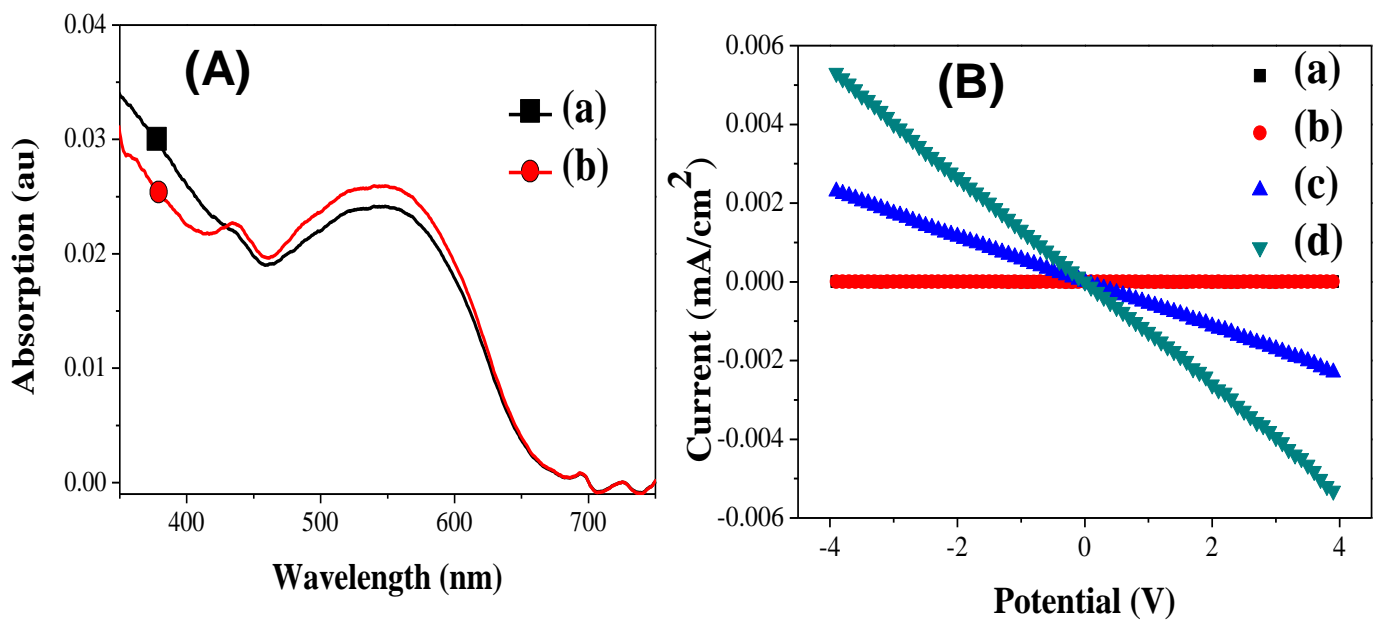

**Figure 3**. (A) UV/Visible absorption spectra of *Co₃(NDC)₃* MOF film on an amine functionalized glass slide using LbL technique: (a) as-prepared, and (b) iodine doped film. (B) Current *vs* applied bias curve of *Co₃(NDC)₃* MOF films on a glass substrate: Doctor-Blade film without doping (a), LbL film without doping (b), Doctor-Blade film with iodine doping (c), and LbL film with iodine doping (d).

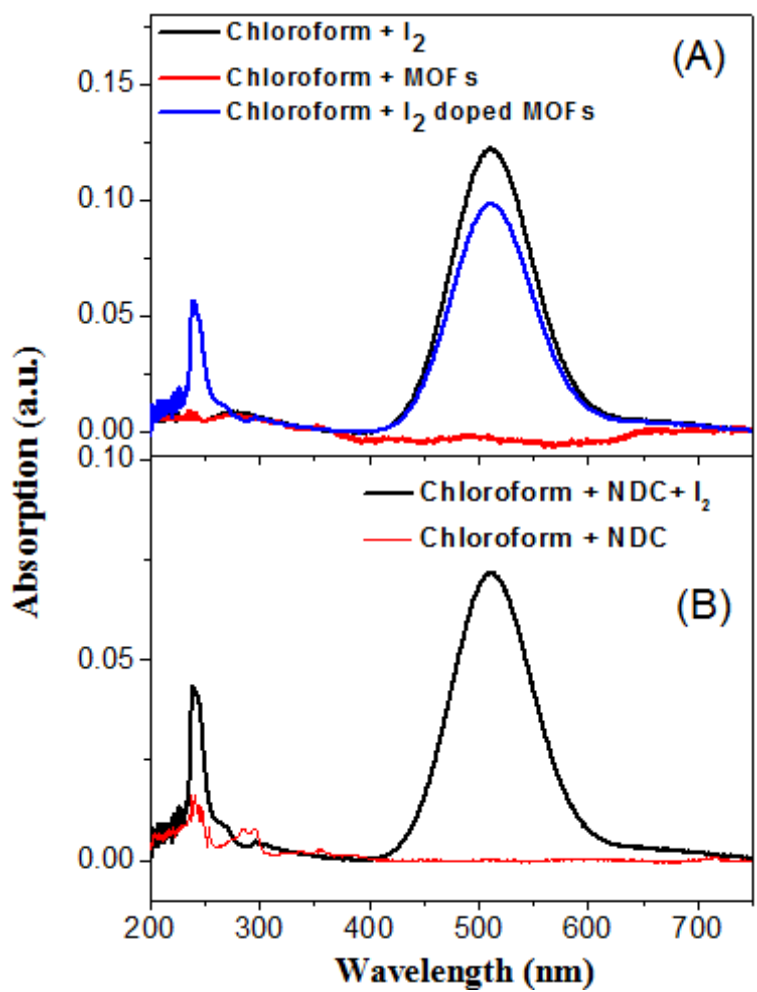

**Figure 4.** UV/Visible absorption spectra of (A) iodine and saturated suspension of undoped and iodine doped $Co_3(NDC)_3$ MOFs in chloroform, and (B) iodine and saturated suspension napthalene dicarboxylic acid (NDC) in chloroform.

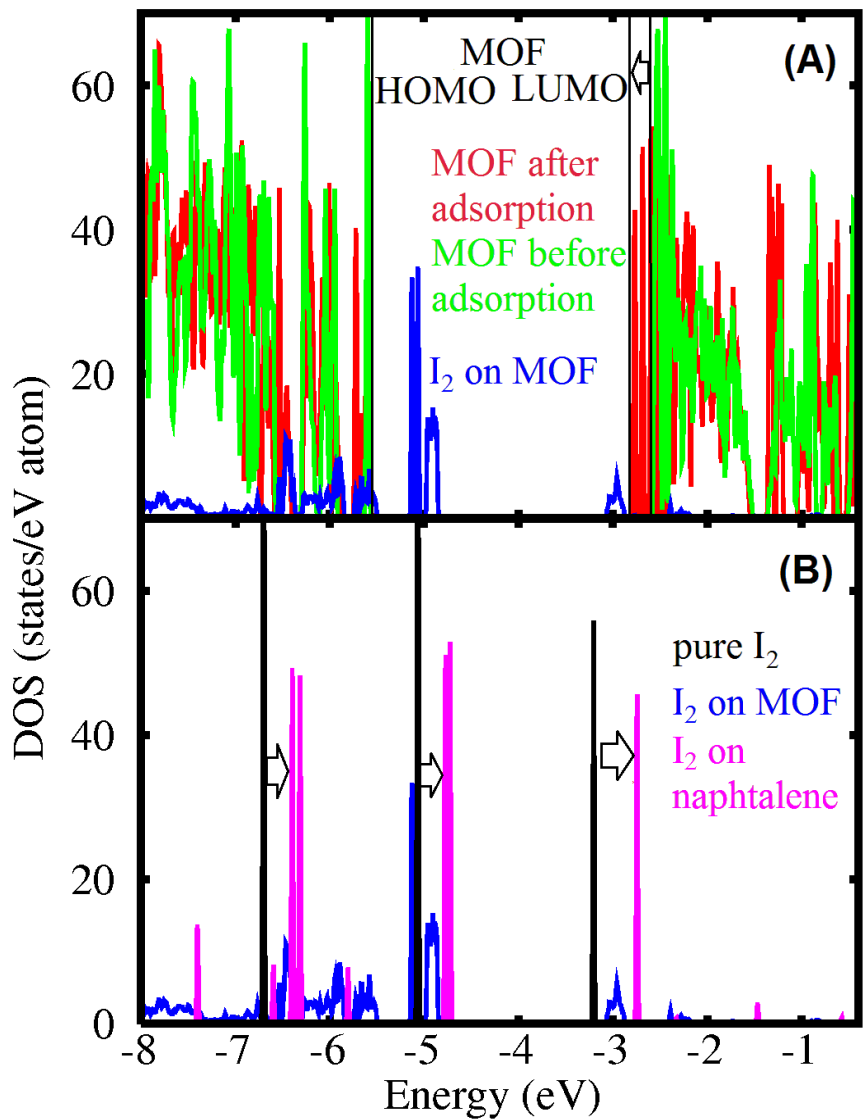

**Figure 5.** Total densities of states of $Co_3(NDC)_3$ MOFs with and without adsorbed iodine molecule (a), and iodine molecule adsorbed on $Co_3(NDC)_3$ MOFs, naphthalene dicarboxylic acid molecule and free iodine molecule in empty box (b). Shifts of the positions of selected orbitals are indicated by arrow.

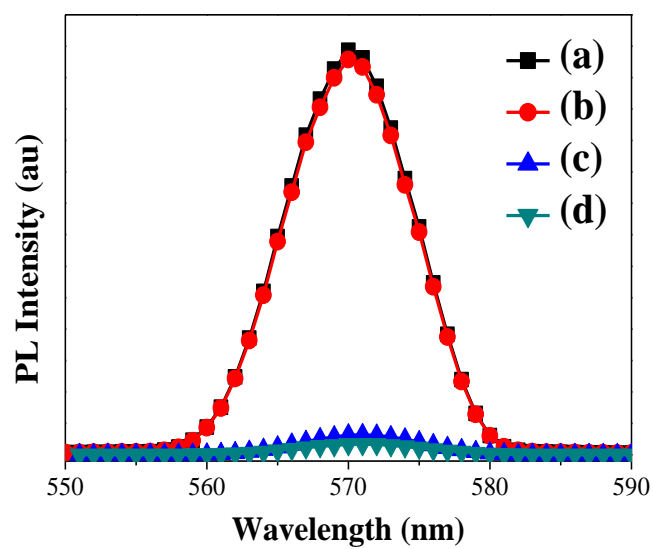

**Figure 6.** PL emission spectrum of (a) Doctor-bladed undoped $Co_3(NDC)_3$ MOF film on ITO substrate, (b) undoped $Co_3(NDC)_3$ MOF LbL film on ITO, (c) Doctor-bladed iodine doped $Co_3(NDC)_3$ MOF film on ITO, and (d) iodine doped $Co_3(NDC)_3$ MOF LbL film on ITO.

# Supporting information

**Charge Transfer Induced Molecular Hole Doping into Thin Film of Metal–Organic-Frameworks**

Deok Yeon Lee,[†] Eun-Kyung Kim,[†] Nabeen K. Shrestha,[†*] Danil W. Bukhvalov,[†*] Joong Kee Lee,[‡] Sung-Hwan Han[†*]

[†]Department of Chemistry, Hanyang University, Seoul 133-791, Republic of Korea

[‡]Korea Institute of Science and Technology, Seoul 136-791, Republic of Korea

*Corresponding:

nabeenkshrestha@hotmail.com; danil@hanyang.ac.kr; shhan@hanyang.ac.kr

# Experimental details

**1.1. Synthesis of bulk *Co$_3$(NDC)$_3$*:** Bulk *Co$_3$(NDC)$_3$* was synthesized using the previously reported method[1] with slightly modified concentrations of metal ion and ligand precursors. In a typical synthesis, an equal volume of 0.344 mmol Co(NO$_3$)$_2$.3H$_2$O in DMF is mixed with 0.115 mmol 2,6-NDC (2,6-naphthalenedicarboxylic acid) dissolved in DMF. The resulting mixture was heated under hydrothermal conditions at 453.15 K for 3 days. All precipitate was filtered out and washed thoroughly with DMF solvent and dried at 50 °C for 3 h. These bulk MOFs were used as the reference material to characterize the MOF films prepared by LbL and DB coating techniques.

**1.2. Doctor blade coating of thin *Co$_3$(NDC)$_3$* film:** 0.5 g of the bulk MOFs was grinded in a mixture of 0.05 g PEG, 2 mL ethanol and 1 mL deionized water until the mixture attends a proper viscosity. The blended semi-liquid paste was then doctor bladed on nonconducting glass and ITO conducting glass substrates. The doctor bladed film was annealed at 250° C for 1 h to decompose the polyethylene glycol (PEG) used as an organic binder in the film.

**1.3. Layer-by-layer (LBL) growth of *Co$_3$(NDC)$_3$* thin film:** MOF layers were grown on amine functionalized nonconducting glass and ITO (Indium Tin Oxide) glass substrates (provided by LumiNano Co., Ltd, S. Korea) for 20 LbL cycles. Each cycle is consisted of dipping the substrate for 1 h into 0.1 M 2,6-naphthalenedicarboxylic acid in DMF solution maintained at 120°C followed by washing the film by dipping in DMF solution, and subsequently dipping the film into 0.1 M Co(NO$_3$)$_2$.3H$_2$O in DMF solution maintained at 120 °C for another 1 h. Finally, the film was again washed with DMF.

**1.4. Iodine doping.** Iodine doping of the framework films was performed by using a slightly modified procedure as reported earlier.[2] In brief, doping was performed by dipping the framework films into a solution of 0.1 M $I_2$ in acetonitrile for 2 h at 50 °C.

**2.1. Characterization of MOF film:** The surface and cross-sectional morphology of the films were investigated using a field enhanced scanning electron microscope (FE-SEM, Hitachi S-4200). The crystal structural of the films was investigated using an X-ray diffractometer (XRD, Rigaku & D/MAX-2500H). The composition of the MOF films was determined using an energy dispersive x-ray analyzer (HORIBA) coupled to the Hitachi S-4200 FE-SEM device and an X-ray photoelectron spectrometer (VG Multilab ESCA 2000 system). Optical property of the films was investigated using an UV/Visible absorption spectrophotometer (VARIAN & CARY-100 Conc). Energy gap and HOMO-LUMO positions were determined from UV/Visible absorption spectra and cyclic voltammogram. Charge transfer phenomenon was studied using a photoluminescence spectroscopy (PL, ILC Technology, Inc & ISS PC-1).

For conductivity measurement, MOF films were deposited on a non-conducting glass substrate, and the contact points were prepared with the silver paste. Conductivity was measured between two points using a computer controlled digital source meter (Keithley-2400). Similarly, conductivity as well as charge carrier concentration, mobility and other semiconducting properties were measured using a Hall Effect measuring instrument (HMS-300: ECOPIA) at 25 °C.

**SI-1. Details about *ab initio* calculations based on density functional theory**

To confirm the experimental observation on the possibility of $I_2$ adsorption on $Co_3(NDC)_3$ MOFs and naphthalene dicarboxylic acid molecule by charge transfer complex formation, and to estimate the charge transfer quantitatively, were performed *ab initio* calculations based on density functional theory (DFT). We performed the calculations for the adsorption of single iodine molecule at various possible sites of realistic $Co_3(NDC)_3$ MOF, and calculate the change of Mulliken populations of iodine, electronic structure and binding energies for the configuration with lowest total energy. We used the pseudo-potential code SIESTA.[3] All calculations were performed using the local density approximation (LDA)[4], which is feasible to model adsorption of molecules with charge transfer.[5,6] Other alternative approach for the modeling of weakly bonded layered systems is GGA+vdW [7-9] is proper only for the systems without charge transfer and lack in description of systems with charge transfer.[10] Crystal structure parameters are taken from the Ref. [1]. For the correct description of electronic structure of $Co_3(NDC)_3$ MOF we employ LDA+U approximation[11] with Coulomb parameter U = 4 eV. This value is rather good for description of molecular systems with transitional metals in oxygen environment (see Ref. [12] and references therein). The atomic positions were fully optimized. The wave functions were expanded with a triple-ζ plus polarization basis of localized orbitals for iodine, double-ζ plus for cobalt, carbon, nitrogen and oxygen and double-ζ for hydrogen. The force and total energy was optimized with an accuracy of 0.04 eV/Å and 1 meV, respectively. All calculations were performed with an energy mesh cut-off of 360 Ry and a **k**-point mesh of 6×6×6 in the Monkhorst-Pack scheme.[13] The binding energies are defined as energy difference between total energy of the systems with adsorbed iodine molecule and with iodine molecule remote from adsorption site. For the calculations of electronic structure of iodine molecule we performed the calculations for single molecule in empty box size of 10×10×10 Å that guaranteed absence of any interactions between molecules.

**Table S1.** Various parameters obtained from Hall Effect measurement of Doctor-blade $Co_3(NDC)_3$ film on a glass substrate.

| | | | |
|---|---|---|---|
| Bulk concentration | $5.12 \times 10^{11}$ (cm$^3$) | Sheet Concentration | $5.12 \times 10^7$ (cm$^2$) |
| Mobility | 1.24 (cm$^2$/Vs) | Conductivity | $1.02 \times 10^{-7}$ (s/cm$^{-1}$) |
| Resistivity | $9.82 \times 10^6$ (Ωcm) | Hall Coefficient | $1.22 \times 10^7$ (m$^2$c$^{-1}$) |

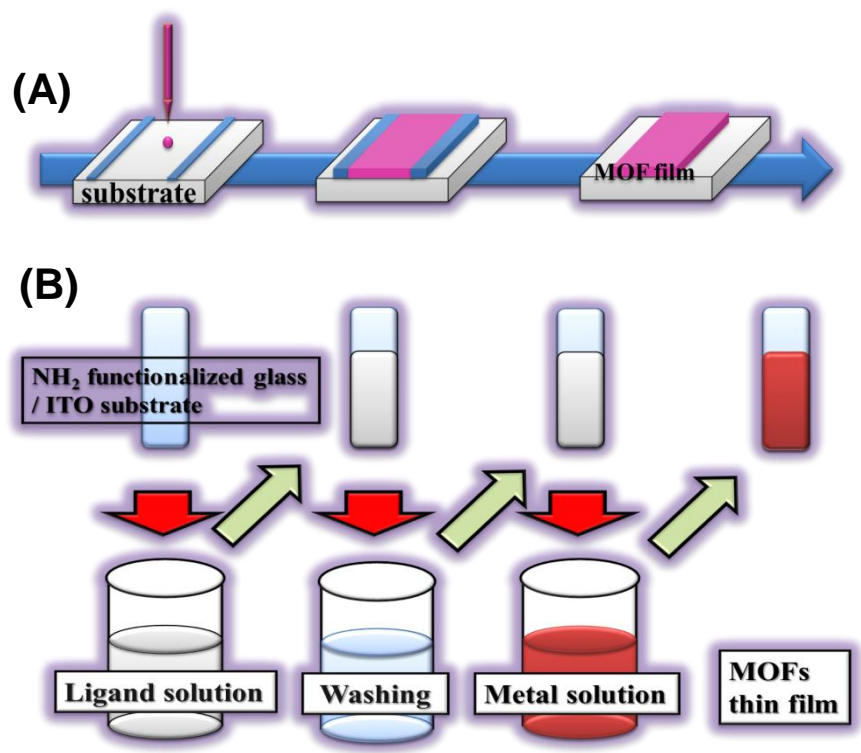

**Scheme S1.** Fabrication of thin film $Co_3(NDC)_3$ using Doctor-Blade coating (A), and Layer-by-Layer (B) techniques.

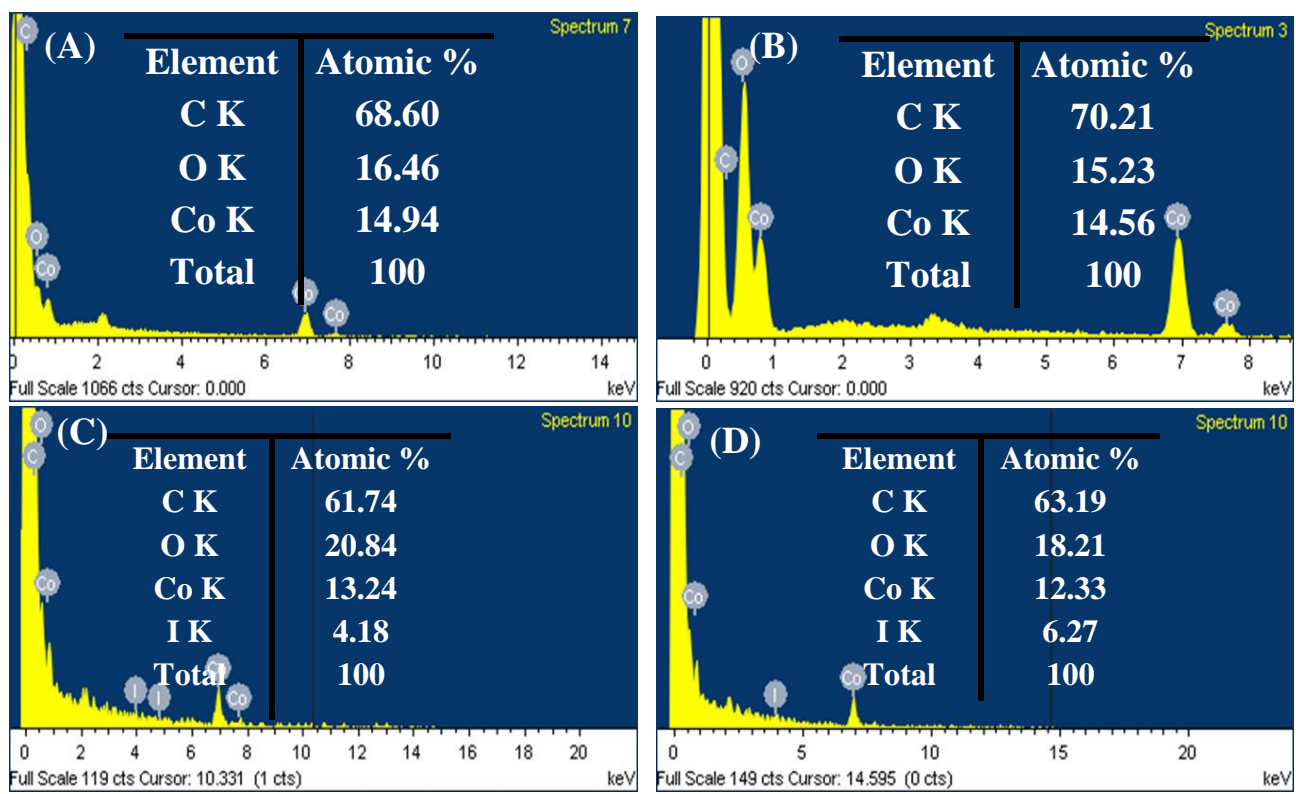

**Figure S1**. EDX analysis for composition of $Co_3(NDC)_3$ MOF film: (A) doctor blade film without doping, and (B) LBL film without doping, (C) doctor blade film with iodine doping, and (D) LBL film with iodine doping.

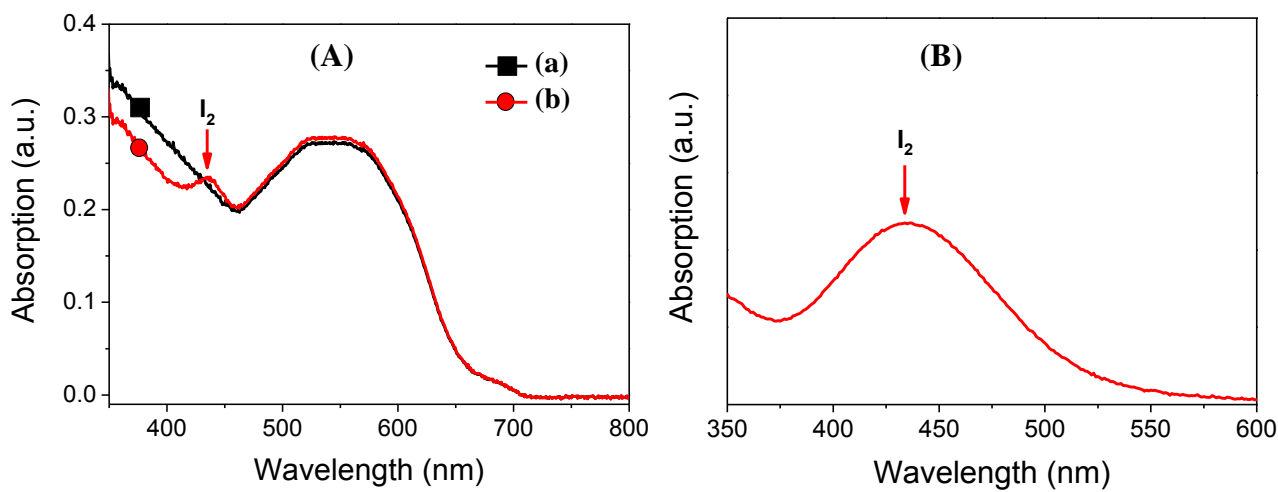

**Figure S2**. (A) UV/Visible absorption spectra of *Co₃(NDC)₃* film on a glass slide using doctor blade coating technique: (a) undoped *Co₃(NDC)₃* film, (b) iodine doped *Co₃(NDC)₃* film. (B) UV/Visible absorption spectra of iodine in acetonitrile solution. Peaks shown by red arrow in Fig. (A) and (B) are at the same position, which suggests that the peak shown by arrow in (A) is due to the presence of iodine.

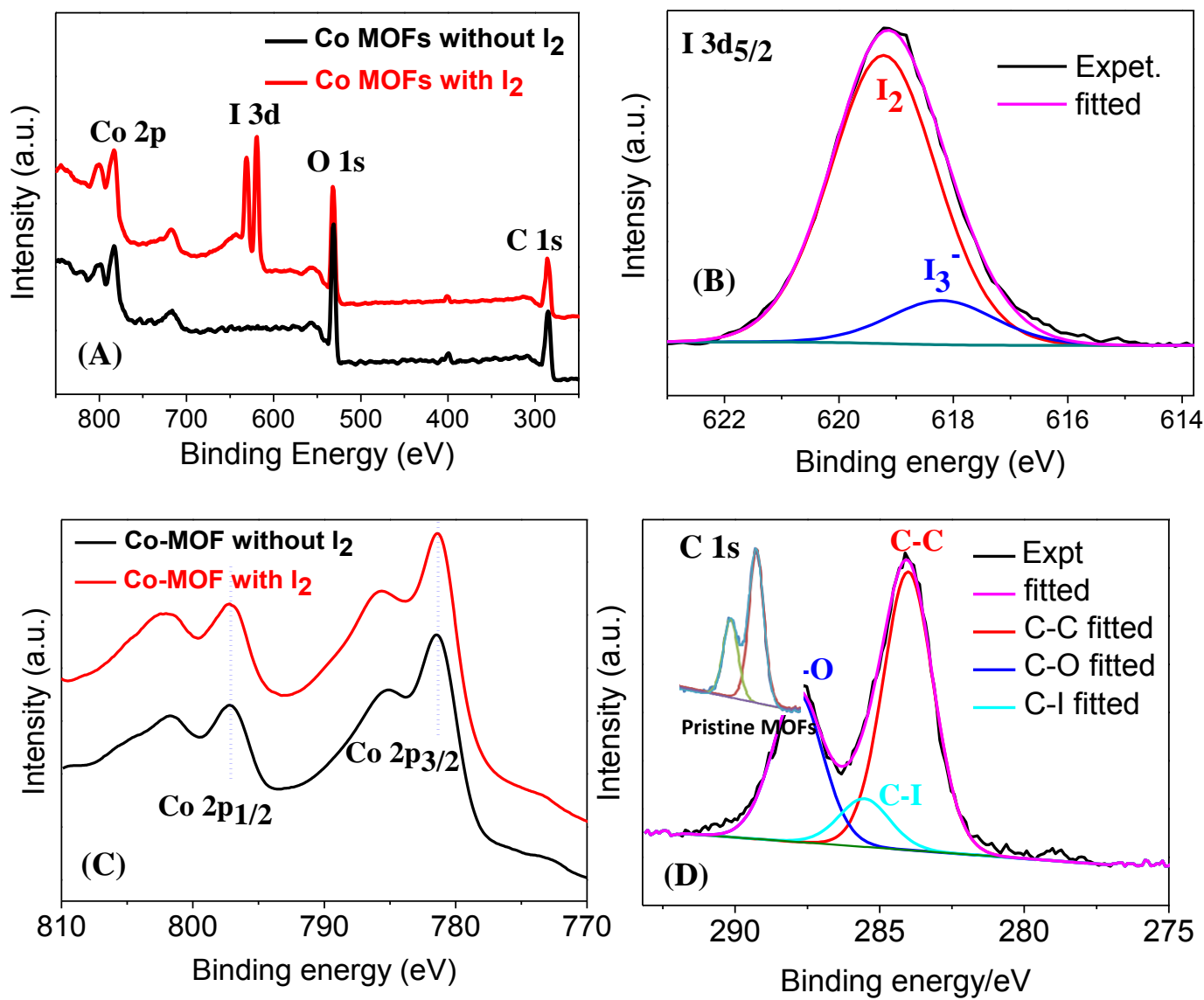

**Figure S3.** XPS spectra of *Co₃(NDC)₃* frameworks with and without I$_2$ doping. (A) Survey spectra, (B) high resolution I 3d$_{5/2}$ spectrum, (C) high resolution Co 2p spectra, and (D) high resolution C 1s spectra of iodine doped frameworks. Inset in Figure D shows the C 1s spectra of pristine frameworks.

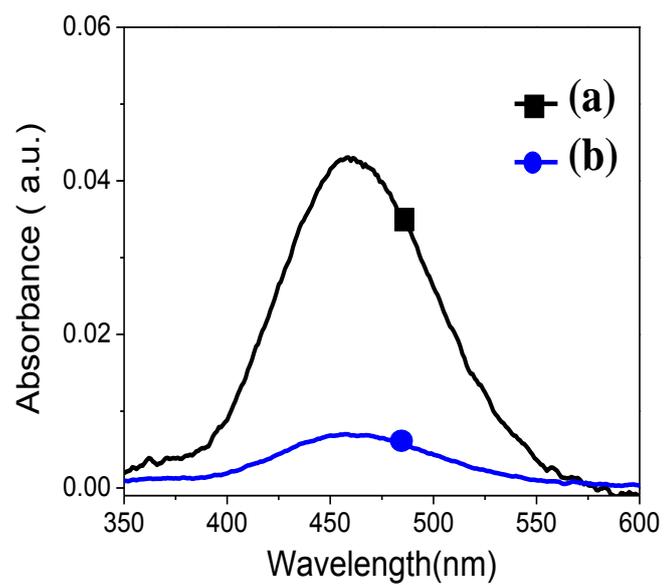

**Figure S4.** UV/Visible absorption spectra of 100 times diluted solution of 0.1 M iodine in acetonitrile solution before/after dipping *Co$_3$(NDC)$_3$* LBL film for 2 h at 50°C. (a) Before, and (b) after dipping.

(A)

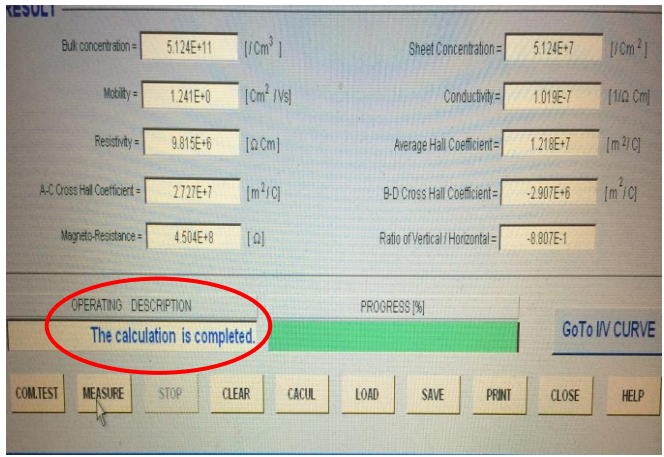

(B)

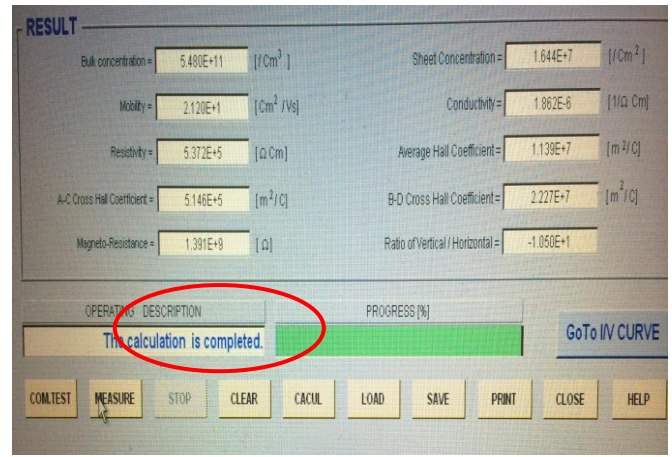

(C)

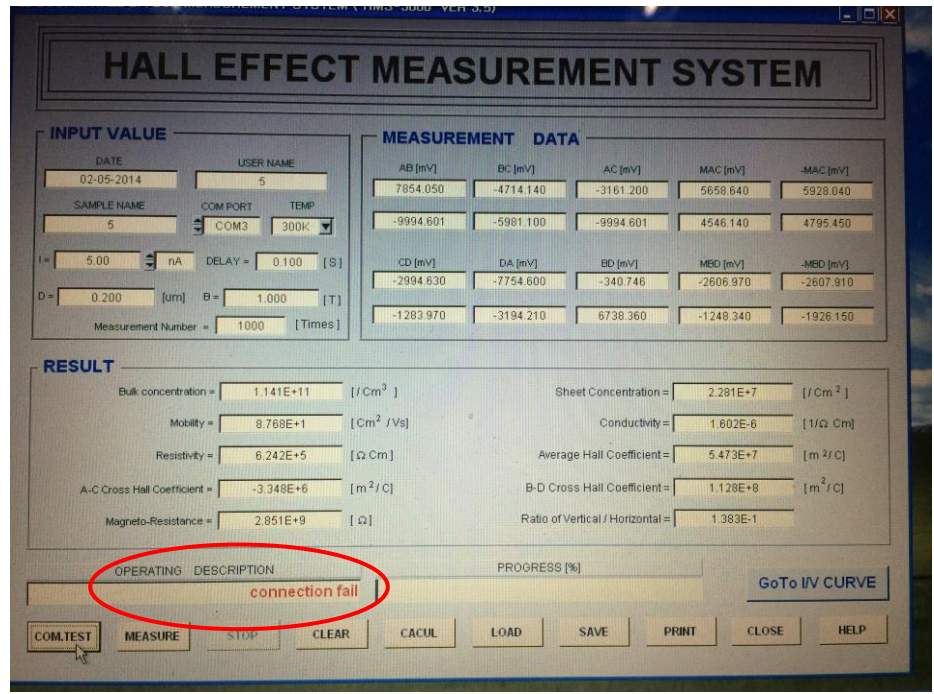

**Figure S5.** Photo images of the monitor of a Hall Effect measuring device displaying the different measured parameters. (A) Iodine doped *Co$_3$(NDC)$_3$* doctor blade film, (B) iodine doped *Co$_3$(NDC)$_3$* LBL film, and (C) undoped *Co$_3$(NDC)$_3$* doctor blade film. Red circled messages show the status of the operation, which indicates if the measurements were successful.

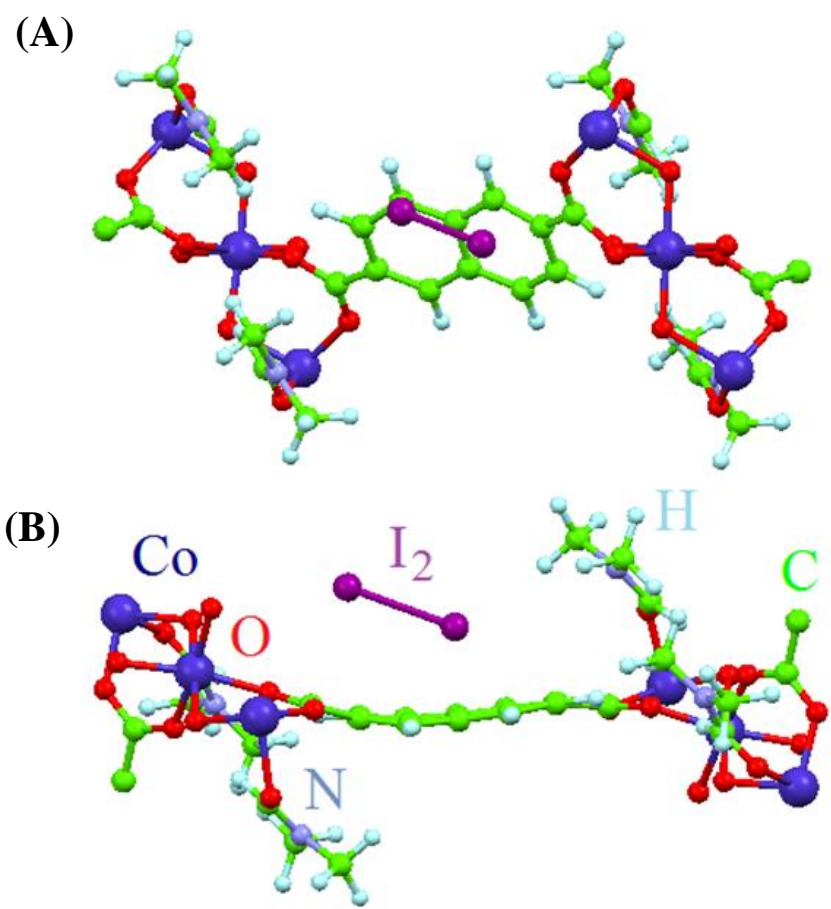

**Figure S6.** Top (a) and side (b) view of optimized atomic structure of $Co_3(NDC)_3$ MOF in vicinity of adsorbed iodine molecule. Note: other part of $Co_3(NDC)_3$ MOF is omitted on the figure for clarity.

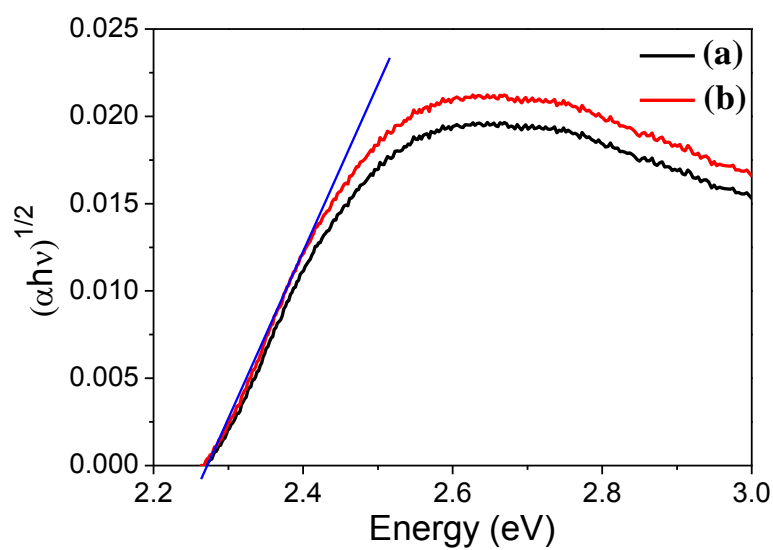

**Figure S7.** Band gap estimation of iodine doped *Co₃(NDC)₃* MOF films. (a) Layer-by-layer film, and (b) doctor blade coating film.

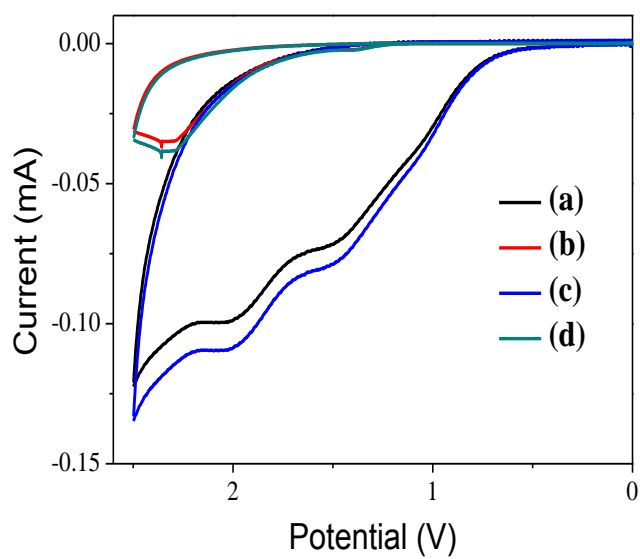

**Figure S8.** Cyclic voltammograms of *Co₃(NDC)₃* LbL films and doctor blade film on an ITO glass substrate under Ar atmosphere. (a) Iodine doped *Co₃(NDC)₃* LbL film, (b) undoped *Co₃(NDC)₃* LbL film, (c) iodine doped *Co₃(NDC)₃* doctor blade film, and (d) undoped *Co₃(NDC)₃* doctor blade film. W.E: MOF film on ITO, C.E: Pt wire, and R.E.: Ag/AgCl; Electrolyte: 0.1 M tetra-n-butylammonium tetrafluoroborate in acetonitrile.

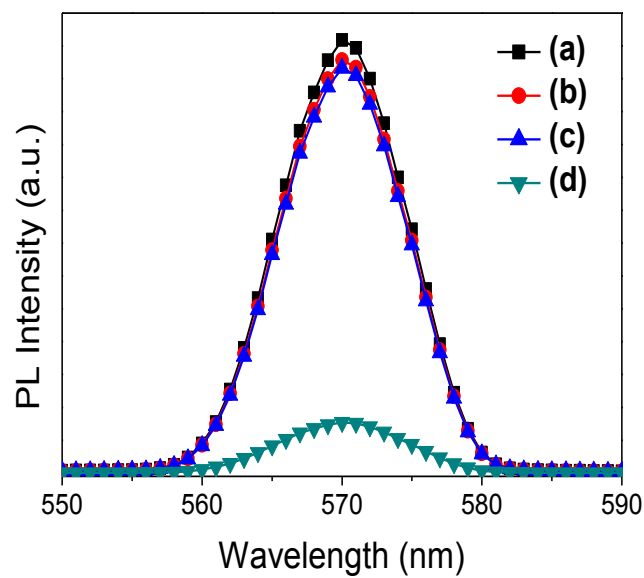

**Figure S9.** PL emission spectra. (a) Undoped *Co$_3$(NDC)$_3$* doctor-blade film on a glass substrate, (b) iodine doped *Co$_3$(NDC)$_3$* doctor-blade film on a glass substrate, (c) undoped *Co$_3$(NDC)$_3$* doctor-blade film on a ITO substrate, (d) iodine doped *Co$_3$(NDC)$_3$* doctor-blade film on an ITO substrate.

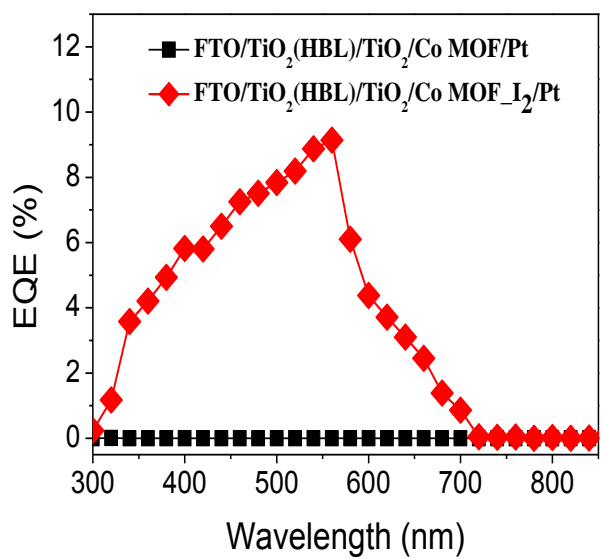

**Figure S10.** Incident Photon to Current Efficiency (IPCE) of a $TiO_2$-based solar cell with the light harvesting $Co_3(NDC)_3$ LBL layer.

## References

[1] Liu, B.; Zou, R. Q.; Zhong, R. Q.; Han, S.; Shioyama, H.; Yamada, T.; Maruta, G.; Takeda, S.; Xu, Q. *Microporous Mesoporous Mater.* **2008**, *111*, 470-477.

[2] Zeng, M. H.; Wang, Q. X.; Tan, Y. X.; Hu, S.; Zhao, H. X.; Long, L. S.; Kurmoo, M. *J. Am. Chem. Soc.* **2010**, *132*, 2561-2563.

[3] Soler, J. M.; Artacho, E.; Gale, J. D.; Garsia, A.; Junquera, J.; Orejon, P.; Sanchez-Portal, D. *J. Phys.: Condens. Matter* **2002**, *14*, 2745-2779.

[4] Perdew, J. P.; Zunger, A. *Phys. Rev. B* **1981**, *23*, 5048-5079.

[5] Wehling, T. O.; Novoselov, K. S.; Morozov, S. V.; Vdovin, E. E.; Katsnelson, M. I.; Geim, A. K.; Lichtenstein, A. I. *Nano Lett.* **2008**, *8*, 173-177.

[6] Boukhvalov, D. W. *Surf. Sci.* **2010**, *604*, 2190-2193.

[7] Zaimbras, E.; Kleis, J.; Schröder, E.; Hyldgaard, P. *Phys. Rev. B.* **2007**, *76*, 155425-155434.

[8] Bučko, T.; Hafner, J.; Lebègue, S.; Ángyán, J. G. *J. Phys. Chem. A.* **2010**, *114*, 11814-11824.

[9] Román-Pérez, G.; Soler, J. M. *Phys. Rev. Lett.* **2009**, *103*, 096102-096105.

[10] Boukhvalov, D. W.; Gornostyrev, Y. N.; Uimin, M. A.; Korolev, A. V.; Yermakov, A. Y. *RSC Adv.* **2015**, *5*, 9173-9179.

[11] Anisimov, V. I.; Aryasetiawan, F.; Lichtenstein, A. I. *J. Phys.: Condens. Matter* **1997**, *9*, 767-808.

[12] Boukhvalov, D. W.; Dobrovitski, V. V.; Kögerler, P.; Al-Saqer, M.; Katsnelson, M. I.; Lichtenstein, A. I.; Harmon, B. N. *Inorg. Chem.* **2010**, *49*, 10902-10906.

[13] Monkhorst, H. J.; Pack, J. D. *Phys. Rev. B* **1976**, *13*, 5188-5192.